\documentclass[fleqn,usenatbib]{mnras}

\usepackage{ae, aecompl, amsmath, amssymb}
\usepackage{bm}           
\usepackage{cancel, mathptmx}
\usepackage{color}
\usepackage{dcolumn}  
\usepackage{epsfig, epsf, etoolbox}
\usepackage{fancyhdr}
\usepackage{graphicx}
\usepackage[T1]{fontenc}
\usepackage{ifthen}
\usepackage{listings}
\usepackage{longtable}
\usepackage{multirow}
\usepackage{newtxtext, newtxmath}
\usepackage{pifont}
\usepackage{subfigure}
\usepackage{tabu}

\usepackage{verbatim}

\usepackage{xcolor}
\usepackage{threeparttable}

\usepackage{graphicx}	
\usepackage{tikz}

\usepackage{arydshln}

\def\nat{Nat} \def\apjl{ApJ~Lett.} \def\apj{ApJ}
\def\apjs{ApJS} \def\aj{AJ} \def\mnras{MNRAS}

\def\pasp{PASP} \def\aap{Astron.~\&~Astrophys.} \def\araa{ARA\&A}
\def\jcap{\ref@jnl{J. Cosmology Astropart. Phys.}}

\def \gsim { \lower .75ex \hbox{$\sim$} \llap{\raise .27ex \hbox{$>$}} }
\def \lsim { \lower .75ex \hbox{$\sim$} \llap{\raise .27ex \hbox{$<$}} }

\newcommand{\lya}{Ly$\alpha$\ }

\newcommand{\ciii}{C\,{\sc iii}]\ }
\newcommand{\civ}{C\,{\sc iv}}

\newcommand{\mgii}{Mg\,{\sc ii}\ }

\newcommand{\oiii}{[O\,{\sc iii}]\ }
\newcommand{\heii}{He\,{\sc ii}\ }

\newcommand{\nv}{N\,{\sc v}\ }

\newcommand{\cv}{\mbox{C\,{\sc v}}}

\definecolor{dkgreen}{rgb}{0,0.6,0}
\definecolor{gray}{rgb}{0.5,0.5,0.5}
\definecolor{mauve}{rgb}{0.58,0,0.82}

\title[High-redshift CLQs]{The first high-redshift changing-look quasars}

\author[Ross {\it et al.}]
{Nicholas~P.~Ross$^{1}$\thanks{E-mail: npross@roe.ac.uk},
Matthew J. Graham$^{2}$,
Giorgio Calderone$^{3}$, 
K. E. Saavik Ford$^{4,5,6}$,  
\newauthor Barry McKernan$^{4,5,6}$ and Daniel Stern$^{7}$  
\\
$^{1}$Institute for Astronomy, University of Edinburgh, Royal Observatory, Blackford Hill, Edinburgh EH9 3HJ, United Kingdom \\
$^{2}$Cahill Center for Astronomy and Astrophysics, California Institute of Technology, Mail Code 249/17, 1200 E California Blvd, Pasadena CA 91125, USA\\
$^{3}$INAF -- Osservatorio Astronomico di Trieste, Via Tiepolo 11, I-34143 Trieste, Italy \\
$^{4}$Department of Science, BMCC, City University of New York, New York, NY 10007, USA \\
$^{5}$Department of Astrophysics, Rose Center for Earth and Space, American Museum of Natural History, Central Park West at 79th Street, NY 10024, USA \\
$^{6}$Graduate Center, City University of New York, 365 5th Avenue, New York, NY 10016, USA\\
$^{7}$Jet Propulsion Laboratory, California Institute of Technology, 4800 Oak Grove Drive, Mail Stop 169-221, Pasadena, CA 91109, USA \\
}

\date{Accepted XXX. Received YYY; in original form ZZZ}

\begin{document}
\label{firstpage}
\pagerange{\pageref{firstpage}--\pageref{lastpage}}
\maketitle

\begin{abstract}
We report on three redshift $z>2$ quasars with dramatic changes in
their \civ\ emission lines, the first sample of changing-look quasars
(CLQs) at high redshift.  This is also the first time the
changing-look behaviour has been seen in a high-ionisation emission
line.  SDSS J1205+3422, J1638+2827, and J2228+2201 show interesting
behaviour in their observed optical light curves, and subsequent
spectroscopy shows significant changes in the \civ\ broad emission
line, with both line collapse and emergence being displayed on
rest-frame timescales of $\sim$240-1640 days.  These are rapid
changes, especially when considering virial black hole mass estimates
of $M_{\rm BH} > 10^{9} M_{\odot}$ for all three quasars.  Continuum
and emission line measurements from the three quasars show changes in
the continuum-equivalent width plane with the CLQs seen to be on the
edge of the full population distribution, and showing indications of
an intrinsic Baldwin effect. We put these observations in context with
recent state-change models, and note that even in their observed
low-state, the \civ\ CLQs are generally above $\sim$5\% in Eddington
luminosity. 
\end{abstract}


\section{Introduction}
Luminous AGN, i.e. quasars, are now seen to significantly vary their
energy output on timescales as short as weeks to months.  This
observation, and the subsequent mismatch in the expected viscous
timescale, which for a $10^{7} M_{\odot}$ central supermassive black
hole (SMBH) is $\sim$hundreds of years, was noted over 30 years ago
\citep[e.g.][]{Alloin1985}. However, with new photometric light-curve
and repeat spectroscopic data, the desire for a deeper understanding
of AGN accretion disk physics has recently re-invigorated the field
\citep[e.g.,][]{Antonucci2018, Lawrence2018, Ross2018, Stern2018}.

The optical continuum variability of quasars was recognised since
their first optical identifications
\citep{MatthewsSandage1963}. 
in the broad emission lines (BELs) of quasars was only recently first
identified \citep{LaMassa2015}.  Samples of over 100 ``changing-look''
quasars (CLQs) or ``changing-state'' quasars (CSQs) have now been
assembled \citep[e.g.,][]{MacLeod2019, Graham2019b}. The community
uses both these terms as a cover for the underlying physics. For sake
of argument, CLQs can potentially be thought of as the extension to
the BELs of quasar continuum variability \citep[e.g., ][]{MacLeod2012}
whereas the CSQs have a state-transition, perhaps similar to that seen
in Galactic X-ray binaries \citep[][]{NodaDone2018, Ruan2019a}. In
this paper, we use the term ``changing-look'', as we are currently
agnostic, and confessedly ignorant, to the underlying physical
processes.

\begin{table*}
  \begin{centering}
    \begin{tabular}{l r  r r   lll lll  r r}
      \hline  \hline 
      Line                 & $\lambda$ &  Transition  & Ionisation   &  \multicolumn{6}{c}{Transition Levels}                                                                                           & Wavenumber   & $A_{i,j}$                    \\
                              &  / \AA\    & Energy / eV &  Energy / eV  &    \multicolumn{3}{l}{Lower}    &  \multicolumn{3}{l}{Upper}                                                         & / cm$^{-1}$    & / $10^{8}$ s$^{-1}$               \\      \hline
      H LyLim           &   912.324   & 13.5984    & 13.5984        & 1$s$                    & $^2$S     & 1/2          & $\infty$                  &                                 &             & 109 678.7       & 1.23$\times10^{-6}$  \\
      H Ly$\alpha$  &  1215.670  & 10.1988    & 13.5984       & 1$s$                     & $^2$S      & 1/2          & 2                             &                                 &             &  82 259.2       &  4.67  \\
      \nv                  &  1238.821  & 10.0082    &  97.8901      &  1$s^{2}$2$s$      &  $^{2}$S   &  1/2         & 1$s^{2}$2$p$          &  $^{2}$P$^{{\rm o}}$ &   3/2    &  80 721.9        & 3.40   \\
      \nv                  &  1242.804  &  9.9762     &  97.8901      &  1$s^{2}$2$s$       &  $^{2}$S   &  1/2        &  1$s^{2}$2$p$         &  $^{2}$P$^{{\rm o}}$ &  1/2     &  80 463.2        & 3.37 \\
      \civ                 &  1548.187  &  8.0083     &  64.4935      &  1$s^{2}$2$s$       &   $^{2}$S  & 1/2         & 1$s^{2}$2$p$          &  $^{2}$P$^{{\rm o}}$ &  3/2     &  64 591.7        & 2.65  \\
      \civ                 &  1550.772  &  7.9950     &  64.4935      & 1$s^{2}$2$s$        &  $^{2}$S   & 1/2         & 1$s^{2}$2$p$          &  $^{2}$P$^{{\rm o}}$ &  1/2     &   64 484.0       & 2.64  \\
      \heii               &   1640.474  &  7.5578     & 54.4178       & 2$p$ 	      &  $^{2}$P$^{{\rm o}}$ &  3/2  &  3$d$ 	                 & $^2$D                  &  5/2        &  60 958.0       & 10.35 \\
      \heii               &   1640.490  &  7.5578     & 54.4178       & 2$p$ 	      &  $^{2}$P$^{{\rm o}}$ &  3/2  &  3$d$ 	                 & $^2$D                  &  3/2        &  60 957.4       & 1.73 \\
      \ciii                 &  1906.683  &  6.5026     & 47.8878       & 1$s^{2}$2$s^{2}$   &   $^{1}$S   & 0            & 1$s^{2}$2$s$2$p$  &  $^{3}$P$^{{\rm o}}$ &   2        &   52 447.1       & 5.19$\times10^{-11}$ \\
      \ciii                 &  1908.734  &  6.4956     & 47.8878       & 1$s^{2}$2$s^{2}$   &  $^{1}$S    & 0            & 1$s^{2}$2$s$2$p$  &  $^{3}$P$^{{\rm o}}$ &  1         &  52 390.8        & 1.14$\times10^{-6}$  \\
       \mgii              &  2795.528  &  4.4338     & 15.0353       & 2$p^{6}$3$s$        &  $^{2}$S    & 1/2        & 2$p^{6}$3$p$          &  $^{2}$P$^{{\rm o}}$ &   3/2     &  35 760.9       & 2.60  \\
      \mgii               &  2802.705  &  4.4224     & 15.0353       & 2$p^{6}$3$s$        &  $^{2}$S    & 1/2        & 2$p^{6}$3$p$          &  $^{2}$P$^{{\rm o}}$ &   1/2     &  35 669.3       & 2.57 \\
      H Ba $\beta$   &  4861.333  &  2.5497     & 13.5984       & 2                              &                 &               & 4                             &                                 &              &  20 564.8       & 0.0842  \\
      \hline   
      \hline
    \end{tabular}
    \caption{
      Strong UV/optical spectral emission lines in quasars, and their
      atomic data from the NIST Atomic Spectra Database
      \citep{Kramida2018, Kramida2019}.  The transition energies are
      $E=hc/\lambda$ for the given wavelength. The ionisation energy is the
      energy required to ionise the given species, e.g., 64.49 eV is needed
      to create a \cv\ ion.  Transition level configurations are given in
      standard spectroscopic notation, with $A_{i,j}$ denoting the transition
      probabilities.}
    \label{tab:atomic_lines}
  \end{centering}
\end{table*}

CLQs to date have primarily been defined according to the
(recombination) Balmer emission line properties with particular
attention paid to the H$\beta$ emission line, observed from optical
spectroscopy. Recent works report on discoveries of \mgii\ changing-look
AGN \citep{Guo2019, Homan2019}. However, current CLQ studies have
primarily been at redshifts $z<1$.

While there have been many studies on triply ionised carbon, i.e.,
\civ, these have tended to focus on broad absorption line quasars
\citep[e.g., Table 1 of][]{Hemler2019} or the Baldwin effect
\citep{Baldwin1977, Bian2012, Jensen2016,
Hamann2017}\footnote{As noted in \citet{Rakic2017}, two different
types of Baldwin effect are present in the literature: {\it (i)} the {\it
global} (or {\it ensemble}) Baldwin effect, which is an
anti-correlation between the emission line equivalent width and the
underlying continuum luminosity of {\it single-epoch} observations of
a {\it large number} of AGN, and {\it (ii)} the {\it intrinsic} Baldwin
effect, the same anti-correlation but in an {\it individual, variable}
AGN \citep{PoggePeterson1992}.}.  Dramatic changes in the
collisionally excited broad {\it emission} line of \civ\ --- and
indeed \ciii\ --- have not to this point been reported.

\begin{table*}
  \centering
  \begin{tabular}{l l   r ll  r r l}
    \hline 
    \hline 
    \multirow{2}{*}{Object} & \multirow{2}{*}{Redshift} & $g$-band & \multirow{2}{*}{MJD} & \multirow{3}{*}{Instrument} & Exposure      &   SDSS               & \multirow{2}{*}{Notes} \\
                                         &                                        &   / mag      &                                 &                                             &  time / sec &   Plate-FiberID  & \\
    \hline  
                                       & 2.068                              & 18.27          &  53498                   & SDSS                                     &  8057            & 2089-427             & Plate quality marginal \\
 J120544.7+342252.4     & 2.071                             &                    &  58538                   & DBSP                                     &  1800            &  ---                      &  Poor conditions \\
                                       & 2.071                             &                     &  58693                  & DBSP                                      &  2400            &  ---                      &  Good conditions        \\
                                       &                                       &                     &                               &                                              &                      &                               &                               \\
                                      & 2.185                              & 19.77           &  54553                   & SDSS                                     &   4801            &  2948-614              & Plate quality good \\
J163852.9+282707.7     & 2.186                              &                    &  55832                     & BOSS                                    &   3600            &  5201-178            & Plate quality good \\
                                      & 2.182                              &                     &  58583                     & LRIS                                    &  1800              &  ---                      & \\
                                      &                                         &                    &                                 &                                            &                      &                              &                                 \\
                                      & 2.217                               & 19.97          &  56189                    &  BOSS                                    &  2700            &   6118-720          & Plate quality good \\
 J222818.7+220102.9   & 2.222                               &                    &  56960                     & eBOSS                                    &  4500            &   7582-790          & Plate quality good \\ 
                                     &  2.222                               &                    &  58693                    &  DBSP                                    & 2400             &    ---                        &    \\
    \hline \hline   
  \end{tabular}
  \caption{Details of our spectroscopic observations.  Redshift errors are
    typically $\pm 0.002$.  SDSS, BOSS, and eBOSS spectra have
    resolving power $\mathcal{R} \equiv \lambda / \Delta \lambda \sim 2000$.  DBSP: Double Spectrograph on the Palomar
    200-inch Hale telescope.  LRIS: Low Resolution Imaging Spectrometer on Keck
    I 10-m telescope.} 
  \label{tab:obs_notes}
\end{table*}

Here, we present new results for three quasars which show dramatic
changes in their \civ\ and \ciii\ broad emission line properties and
underlying continuum. These are some of the first examples of CLQs at
high ($z>1$) redshift. Moreover, these are the first cases for
substantial changes of ions with high ionisation potentials (I.P.'s
$>$2 Rydberg), thus linking the ionizing photons to the energetic
inner accretion disk.

Details of the atomic transitions that produce strong rest-frame
UV/optical lines in quasars are given in
Table~\ref{tab:atomic_lines}. In this paper we use the wavelengths of
1548.202 and 1550.774~\AA\ for the \civ\ doublet
\citep{Kramida2018}. For ionisation energies, 47.89 eV (3.519 Ry) is
required for doubly-ionised C\,{\sc iii} to become triply-ionised
\civ, and 64.49 eV (4.74 Ry) is needed to ionise \civ\ itself. This
energy corresponds to a thermal temperature of $T \gtrsim$
4$\times10^{5}$ K, implying a heating energy source of (soft) X-ray
photons.

\citet{Wilhite2006} examine \civ\ variability in a sample of 105
quasars observed at multiple epochs by the Sloan Digital Sky Survey
\citep[SDSS;][]{York2000, Stoughton2002, Abazajian2009}.  They find a
strong correlation between the change in the \civ\ line flux and the
change in the line width, but no correlations between the change in
flux and changes in line center or skewness.  These authors find that
the relation between line flux change and line width change is
consistent with a model in which a broad line base varies with greater
amplitude than the line core. The \civ\ lines in these high-luminosity
quasars appear to be less responsive to continuum variations than
those in lower luminosity AGN.

\citet{Richards2011} explored the broad emission line region in over
30,000 $z > 1.54$ SDSS quasars, concentrating on the properties of the
\civ\ emission line. These authors consider two well-known effects
involving the \civ\ emission line: {\it (i)} the anti-correlation
between the \civ\ equivalent width (EW) and luminosity (i.e., the
Baldwin effect) and {\it (ii)} the blueshifting of the peak of \civ\
emission with respect to the systemic redshift.  We denote the
velocity offset of emission lines as $V_{\rm off}$ and use the
convention that a positive $V_{\rm off}$ value means the line is
blueshifted while a negative $V_{\rm off}$ value means the line is
redshifted.  \citet{Richards2011} find a blueshift of the \civ\
emission line is nearly ubiquitous, with a mean shift of $\langle
V_{\rm off}\rangle_{\rm RQ} \sim$810 km s$^{-1}$ for radio-quiet (RQ)
quasars and $\langle V_{\rm off}\rangle_{\rm RL} \sim$360 km s$^{-1}$
for radio-loud (RL) objects. \citet{Richards2011} also find the
Baldwin effect is present in both their RQ sample and their RL sample.
They conclude that these two \civ\ parameters (EW and blueshift) are
capturing an important trade-off between disk and wind components in
the disk-wind model of accretion disks \citep[e.g.,][]{Murray1995,
Elvis2000, Proga2000, Leighly2004b}, with one dominating over the
other depending on the shape of the quasar spectral energy
distribution (SED).

Using the multi-epoch spectra of 362 quasars from the SDSS
Reverberation Mapping project \citep[SDSS-RM; ][]{Shen2015, Shen2019},
\citet{Sun2018} investigate the blueshift of \civ\ emission relative
to \mgii\ emission, and its dependence on quasar properties. They
confirm that high-blueshift sources tend to have low \civ\ EWs, and
that the low-EW sources span a range of blueshift. Other
high-ionisation lines, such as \heii, also show similar blueshift
properties. The ratio of the line width of \civ\ to that of \mgii\
increases with blueshift. \citet{Sun2018} also find that quasar
variability might slightly enhance the connection between the \civ\
blueshift and EW, though further investigation here is warranted. They
also find that quasars with the largest blueshifts are less variable
and tend to have higher Eddington ratios, though Eddington ratio alone
might be an insufficient condition for the \civ\ blueshift. Recent
investigations also include \citet{Meyer2019} and
\citet{Doan2019}. \citet{Dyer2019} provide a detailed analysis of 340
quasars at high redshift ($1.62<z<3.30$) from the SDSS-RM project,
which we compare with our results in Section~\ref{sec:theory}.

The purpose of this paper is, for the first time, to systematically
access and report on the CLQ phenomenon at high ($z>2$)
redshift. While accessing this phenomenon at an earlier cosmic epoch
is somewhat interesting, the main value of this study is to move from
the low-ionisation energy Balmer emission line series at rest-frame
optical wavelengths to the high-ionisation emission lines, in
particular \civ\ $\lambda$1549, at rest-frame UV wavelengths.

This paper is organised as follows. In Section 2, we describe our
sample selection, catalogues, and observational data sets.  Section 3
presents the high-redshift quasars and reports their time-variable
line properties.  We provide a brief theoretical discussion in Section
4, and Section~5 presents our conclusions.  We report all magnitudes
on the AB zero-point system \citep{Oke_Gunn1983, Fukugita1996} unless
otherwise stated. For the {\it WISE} bands, $m_{\rm AB} = m_{\rm Vega}
+ m$ where $m = 2.699$ and 3.339 for {\it WISE} W1 (3.4$\mu$m) and W2
(4.6$\mu$m), respectively \citep{Cutri2011, Cutri2013}. We adopt a
flat $\Lambda$CDM cosmology with $\Omega_{\Lambda} = 0.73 $,
$\Omega_{\rm M} = 0.27$, and $h = 0.71$. All logarithms are to the
base 10.

\section{CLQ Selection and Line Measurements}
In this section we present the photometric data used to select the
CLQs, and then give details to the multiwavelength data where we have
it. We then give details of the spectroscopic data including emission
lines measurements.

\subsection{Photometry}
\subsubsection{Optical Photometry}
We use optical data from the Catalina Real-time Transient Survey
\citep[CRTS;][]{Drake2009, Mahabal2011}, the Panoramic Survey
Telescope and Rapid Response System \citep[PanSTARRS;][]{Kaiser2010,
Stubbs2010, Tonry2012, Magnier2013}, and the Zwicky Transient Facility
\citep[ZTF;][]{Bellm2019_ZTFOverview}. 

The CRTS archive\footnote{http://catalinadata.org} contains the
Catalina Sky Survey data streams from three telescopes -- the 0.7-m
Catalina Sky Survey (CSS) Schmidt and 1.5-m Mount Lemmon Survey (MLS)
telescopes in Arizona, and the 0.5-m Siding Springs Survey (SSS)
Schmidt in Australia. CRTS covers up to $\sim$2500 deg$^2$ per night,
with four exposures per visit, separated by 10 min. The survey observes
over 21 nights per lunation. The data are broadly calibrated to
Johnson $V$ \citep[for details, see][]{Drake2013} and the current CRTS
data set contains time series for approximately 400 million sources to
$V \sim 20$ above Dec $> -30$ from 2003 to 2016 May (observed with CSS
and MLS) and 100 million sources to $V \sim 19$ in the southern sky
from 2005 to 2013 (from SSS). CRTS has been extensively used to study 
quasar variability \citep[e.g.,][]{Graham2014, Graham2015, Graham2015Nature,
Graham2017, Graham2019b, Stern2017, Stern2018, Ross2018}.

PanSTARRS data is obtained via the Pan-STARRS Catalog Search
interface\footnote{https://catalogs.mast.stsci.edu/panstarrs/}.  We
query the PS1 DR2 Detection catalog.

The Zwicky Transient Facility (ZTF; http://ztf.caltech.edu) is a new
robotic time-domain sky survey capable of visiting the entire visible
sky north of $-$30 declination every night. ZTF observes the sky in
the $g$, $r$, and $i$-bands at different cadences depending on the
scientific program and sky region \citep{Bellm2019_ZTFSurveys,
Graham2019_ZTF}. The ZTF 576 megapixel camera with a 47 deg$^{2}$
field of view, installed on the Samuel Oschin 48-inch Schmidt
telescope at Palomar observatory, can scan more than 3750 deg$^{2}$
per hour, to a 5$\sigma$ detection limit of 20.7 mag in the $r$-band
with a 30-sec exposure during new moon \citep{Masci2019}.
\begin{figure*}
  \centering
  \includegraphics[width=16.7cm, trim=0.3cm 0.05cm 0.45cm 0.1cm, clip]
  {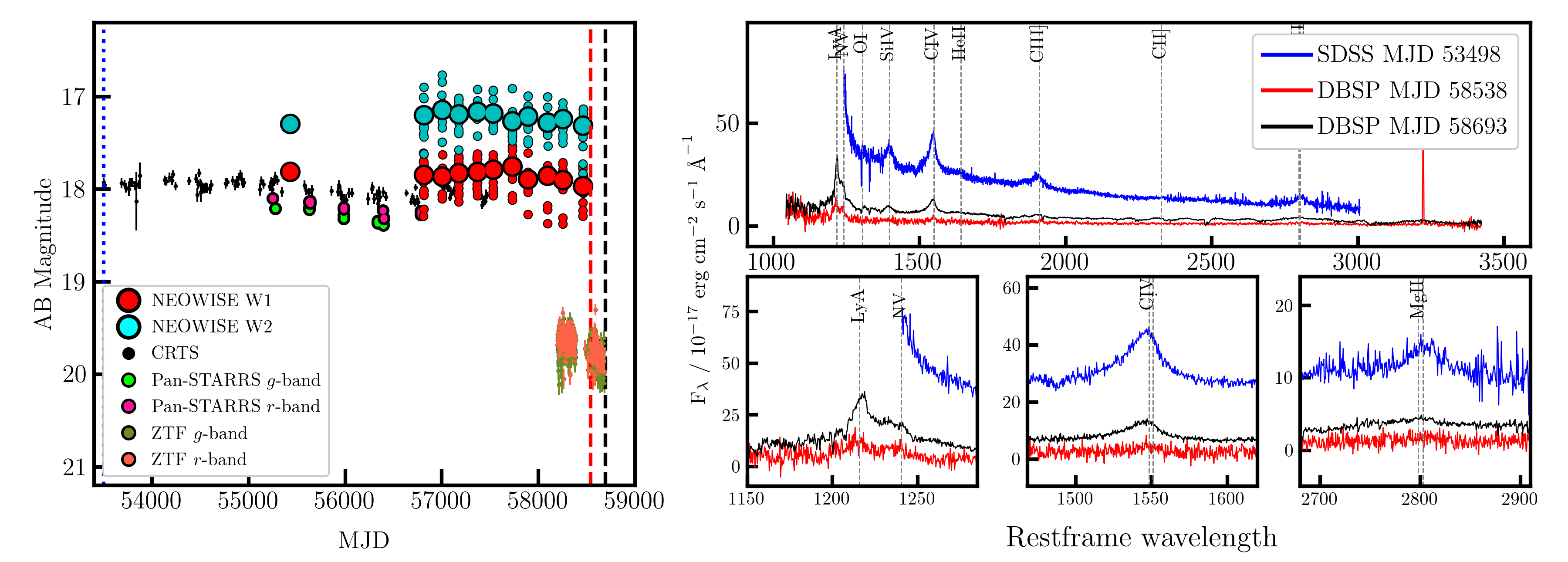}
  \includegraphics[width=16.7cm, trim=0.3cm 0.05cm 0.40cm 0.1cm, clip]
  {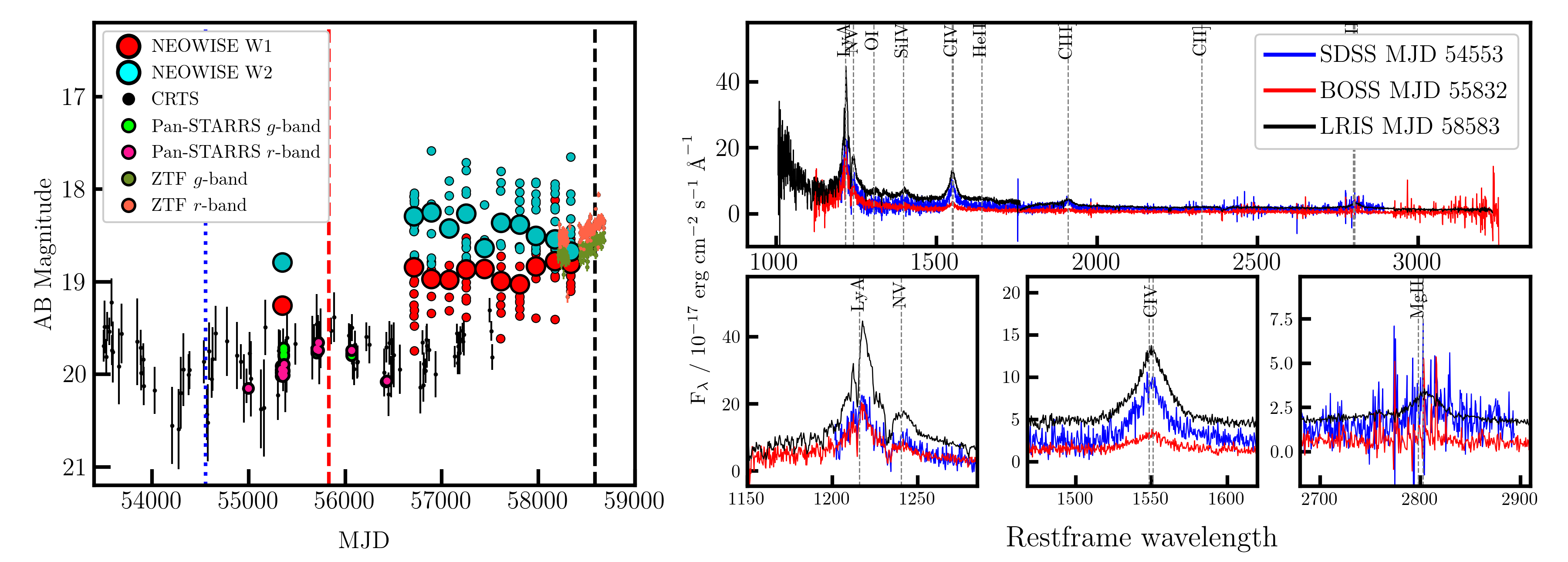}
  \includegraphics[width=16.7cm, trim=0.3cm 0.0cm  0.35cm 0.1cm, clip]
  {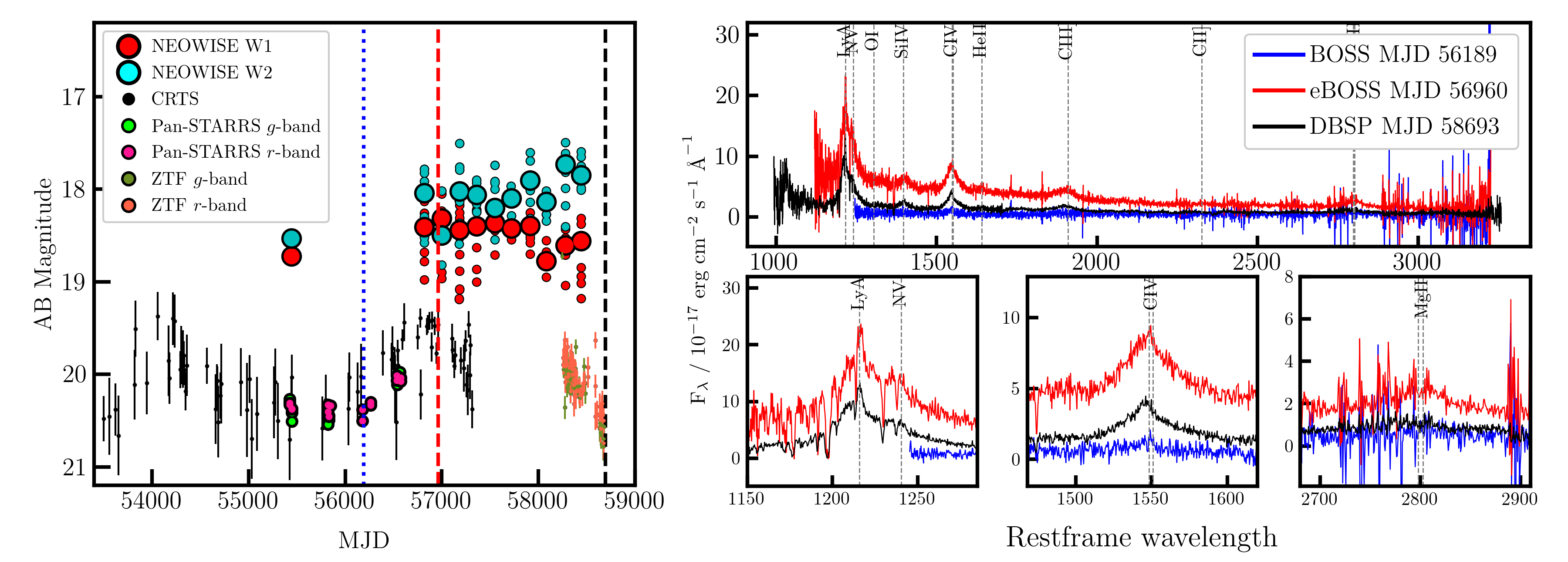}
  \vspace{-12pt}
  \caption[]{The three high-redshift CLQ quasars; 
    J1205+3422 (top), 
    J1638+2827 (middle) and  
    J2228+2201 (bottom). 
    The light curve data are presented in the panels on the left hand side, with the timing of the
    spectral observations indicated with vertical dashed lines. 
    The spectra are on the right hand side, with zoom-in's on the Ly$\alpha$-\nv 
    complex, the \civ\ line, and the \mgii\ line.}
  \label{fig:civ_clqs}
\end{figure*}

\subsubsection{Mid-Infrared Photometry}
Mid-infrared data (3.4 and 4.6$\mu$m) is available from the beginning
of the {\it WISE} mission \citep[2010 January; ][]{Wright2010} through
the fifth-year of {\it NEOWISE-R} operations \citep[2018 December;
][]{Mainzer2011}. The {\it WISE} scan pattern leads to coverage of the
full-sky approximately once every six months (a ``sky pass''), albeit
it with a gab between 2011 February and 2013 October when the
satellite was placed in hibernation. Hence, our light curves have a
cadence of 6 months with a 32-month sampling gap.

\subsection{CLQ Selection}
Our high-redshift CLQs were identified as follows.  We selected all
64,774 SDSS DR15 sources with $z > 0.35$ classified as `quasar', that
had at least two spectra separated by $\geq 100$ days, and that had a
corresponding CRTS light curve. We fitted a damped random walk to the
CRTS data via Gaussian process regression (using the
\href{https://gpy.readthedocs.io/en/deploy/}{GPy Python} library).
This allowed us to predict the photometric magnitudes at the epochs of
the SDSS spectra for each source.  Those with $|\Delta V| > 0.3$ were
then selected for visual inspection, looking for at least a 50\%
change in magnitude between the two epochs.  Only three quasars, SDSS
J120544.7+342252.4 (hereafter J1205+3422), SDSS J163852.93+282707.7
(hereafter J1638+2827) and SDSS J222818.76+220102.9 (hereafter
J2228+2201), satisfied these selection criteria and showed interesting
or dramatic emission line behaviour.

\begin{table*}
  \small
  \begin{centering}
    \begin{tabular}{l  l | c c | c c c c | c}
      \hline
      \hline
              &          &  \multicolumn{2}{c}{Cont. @ 1450\AA }                  &   \multicolumn{4}{c}{CIV 1549\AA}                                                                                & Virial product \\
 Object   & MJD  &      $\nu L_{\nu}$    &         Slope                             &   Luminosity         &     FWHM                 &    V$_{\rm off}$    & EW                                 & log($\nu L_{\nu}^{0.5} \times {\rm FWHM}^2$) \\
              &         & $10^{42}$ erg s$^{-1}$ & (F$_\lambda \propto \lambda^\alpha$)  & $10^{42}$ erg s$^{-1}$ &  km s$^{-1}$   &     km s$^{-1}$   &  \AA                              &  $\log(M/M_{\odot})$ \\
      \hline
                     & 53498   &  38129  $\pm$ 39  &  -1.57 $\pm$ 0.01 &   898 $\pm$ 15    &   6024 $\pm$  120  &  944 $\pm$  33  & 37.78 $\pm$  0.62  &   9.85 $\pm$ 0.02\\
 J1205+3422 & 58538$^*$ & 2725 $\pm$ 40 &  -1.27 $\pm$ 0.05 &  161 $\pm$ 22    &  14997 $\pm$ 2500 &  922 $\pm$ 690  & 91.68 $\pm$ 12.73  &  10.07 $\pm$ 0.13\\
                     & 58693    &  8550 $\pm$ 11    &  -1.41 $\pm$ 0.01 &  385.2 $\pm$ 3.9 &   7109 $\pm$   91   & 1085 $\pm$  29  &   71.32 $\pm$  0.71  &   9.67 $\pm$ 0.01\\  
                      &              &                               &                               &                              &                                 &                             &                                   &                              \\
                      & 54553   & 3579 $\pm$ 49    &  -1.16 $\pm$ 0.07  & 293.6  $\pm$ 8.8 &  4630 $\pm$  180   & 183 $\pm$  57   & 127.67 $\pm$  3.80  &    9.11 $\pm$ 0.04\\  
 J1638+2827 & 55832   & 2340 $\pm$ 15    &  -2.17 $\pm$ 0.05  &  84.6  $\pm$ 4.2  &  4733 $\pm$  290   & 172 $\pm$  89   &  59.61 $\pm$  2.98  &    9.04 $\pm$ 0.05\\  
                      & 58583  & 7793 $\pm$ 19     &  -2.02 $\pm$ 0.01  & 367.5  $\pm$ 4.3 &  4511 $\pm$   71    &  94 $\pm$  24   &  77.89 $\pm$  0.91  &  9.25 $\pm$ 0.01\\  
                      &             &                               &                                &                              &                                 &                            &                                   &                            \\
                     & 56189$^*$ & 607 $\pm$ 49  &  -0.00 $\pm$ 0.16 &  72.8  $\pm$ 11.3 & 14993 $\pm$ 2400 &  828 $\pm$ 691   & 162.89 $\pm$ 25.37  &  9.74 $\pm$ 0.14\\  
 J2228+2201 & 56960   & 7842 $\pm$ 25     &  -1.72 $\pm$ 0.02 & 301.0  $\pm$ 7.2  &  7136 $\pm$  210   & -276 $\pm$  64   &  61.99 $\pm$  1.48  &    9.65 $\pm$ 0.03\\ 
              & 58693   & 2388.4 $\pm$ 6.8 &  -1.22 $\pm$ 0.01 & 145.4  $\pm$ 1.5  &  6084 $\pm$   83    &   168 $\pm$  28   &  94.77 $\pm$  0.99  &    9.26 $\pm$ 0.01\\
      \hline
      \hline
    \end{tabular}
    \caption{Continuum at 1450\AA\ and \civ\ spectral measurements for the
      three quasar considered in this work, at all observation epochs, as
      calculated by QSFit. $^{*}$The \civ\ line is very faint (with respect
      to the continuum), and the associated estimates are likely unreliable.
      For the emission-line velocity offsets, a positive value means the line is blueshifted. 
      The last column shows the virial product calculated as $\nu
      L_{\nu}^{0.5} \times {\rm FWHM}^2$.}
    \label{tab:QSFit-results}
  \end{centering}
\end{table*}

\subsection{Spectroscopy}
An overview of our spectroscopic observations is given in
Table~\ref{tab:obs_notes}.  The archival spectra are from the SDSS
\citep{Stoughton2002, DR7, Schneider2010}, the SDSS-III Baryon
Oscillation Spectroscopic Survey \citep[BOSS; ][]{Eisenstein2011,
Dawson2013, Smee2013, Alam2015, Paris2017} and the SDSS-IV Extended
Baryon Oscillation Spectroscopic Survey \citep[eBOSS; ][]{Dawson2016,
Abolfathi2018, Paris2018}.  These quasars were targetted via a range
of techniques and algorithms \citep[see][]{Richards2002, Ross2012,
Myers2015}. The SDSS, BOSS, and eBOSS data are supplemented by spectra
from the Low Resolution Imaging Spectrometer \citep[LRIS; ][]{Oke1995}
on the 10-m Keck {\sc I} telescope and the Double Spectrograph (DBSP)
instrument on the 200-inch Palomar Hale telescope.

Proper comparison of the spectra requires reliable flux calibrations,
which is a challenge since not all data were taken in photometric
conditions.  SDSS spectra are spectrophotometrically calibrated.  BOSS
and eBOSS spectra have spectrophotometric corrections applied in the
latest data release \citep{Hutchinson2016, Jensen2016, Margala2016}.
However, some of the Palomar and Keck data suffered from variable
conditions and clouds.  For lower redshift AGN, the narrow \oiii
emission line is typically used to align spectrophotometry since it is
spatially extended and not expected to change significantly on human
timescales \citep[e.g.,][]{Barth2011}. However, due to the
high-redshift of our quasars, \oiii is not available to us to use for
scaling the spectra.  Instead, we use photometric data from ZTF since
our Palomar and Keck data were all taken after MJD 57500.  This
provides broad-band photometry at the time of the spectroscopy, which
we use to scale the spectra whose initial calibration is based on
spectrophotometric standards observed on the same nights.

\subsection{Emission Line and Power-law Slope Measurements}
We use the measured quasar emission line properties from several catalogues: 
\citet{Shen2011}, \citet{Hamann2017}, \citet{Kozlowski2017}, and
\citet{Calderone2017}.

In particular, we use the Quasar Spectral Fitting (QSFit) software
package presented in \citet{Calderone2017}. This provides luminosity
estimates as well as width, velocity offset and EW of 20 emission
lines, including \civ, \ciii, and \mgii.  We process and fit all nine
spectra using the lastest version (v1.3.0) of the QSFit
\href{https://qsfit.inaf.it/cat_1.30/onlinefit.php}{online
calculator}. The host galaxy and blended iron emission at rest-frame
optical wavelengths components are automatically disabled when they
can not be constrained by the available data, as is the case for all
our quasars.  Power-law continuum slopes, $\alpha$, where $f_{\lambda}
\propto \lambda^{\alpha}$, are also reported in these catalogues and
from QSFit.

\section{Results}
\subsection{Photometric and Overall Spectral Evolution}
Figure~\ref{fig:civ_clqs} presents the optical and mid-infrared light
curves for three high-redshift CLQ quasars.  Figure~\ref{fig:civ_clqs}
also indicates the spectra for each epoch with vertical dashed lines
in the light curves.

For J1205+3422, our spectral observations cover 5195 observed days,
corresponding to 1691 days in the rest-frame. This quasar was
initially identified in SDSS in 2005 May as a bright, $g \approx
18.0$, blue-sloped quasar with broad Si\,{\sc iv}/O\,{\sc iv}, \civ,
\ciii, and \mgii. \ciii\ and \civ\ have large blueshifts,
$\approx$2600$\pm$150 and $\approx$1150$\pm$100 km s$^{-1}$,
respectively.  By 2019, however, the optical brightness dropped by
nearly 2 magnitudes and the spectra are significantly less blue.  \lya
and \nv are detectable in both 2019 spectra, but is just blueward of
the wavelength range covered by the original SDSS spectrum. The \civ\
and \ciii\ lines have faded significantly between the SDSS observation
in 2005 and the Palomar observations in 2019.  Note that \civ\ is
extremely faint in the 2019 February spectrum, but that night suffered
from poor conditions.

For J1638+2827, our spectral observations cover 4030 observed days,
corresponding to 1265 days in the rest-frame. Here, in the initial
epoch spectrum, \civ\ is broad and bright, as is \ciii. However,
$\approx$400 rest-frame days later, the broad \civ\ and \ciii\
emission lines have faded, the continuum slope around 1400\AA\ has
changed from $\approx-1.48$ to $\approx-2.25$, but the \lya/\nv
emission complex is very similar in shape and line flux
intensity. Around 870 days in the rest-frame after the second spectral
epoch, at the time of the third spectral epoch, the source has
brightened from optical magnitudes of $\sim 20$ mag to $\sim 18.5$
mag. Ly$\alpha$, N\,{\sc v}, \civ, \ciii, and \mgii\ are all apparent
and broad, with \mgii\ being seen for the first time at high
signal-to-noise. An absorption feature between \lya and \nv is seen in
all three spectral epochs.

For J2228+2201, our spectral observations cover 2504 observed days,
corresponding to 778 days in the rest-frame. Over the course of 240
rest-frame days, \civ\ and \ciii\ both {\it emerge} as BELs and the
standard UV/blue continuum slope increases in flux.  Then, over the
course of 538 days in the rest-frame, the broad emission, while still
very present, reduces in line flux the UV/blue continuum diminishes.
However, the third epoch still corresponds to a source more luminous
than the initial BOSS spectrum.

\begin{figure}
  \centering
  \includegraphics[width=8.40cm, trim=0.37cm 0.3cm 0.0cm 0.2cm, clip]{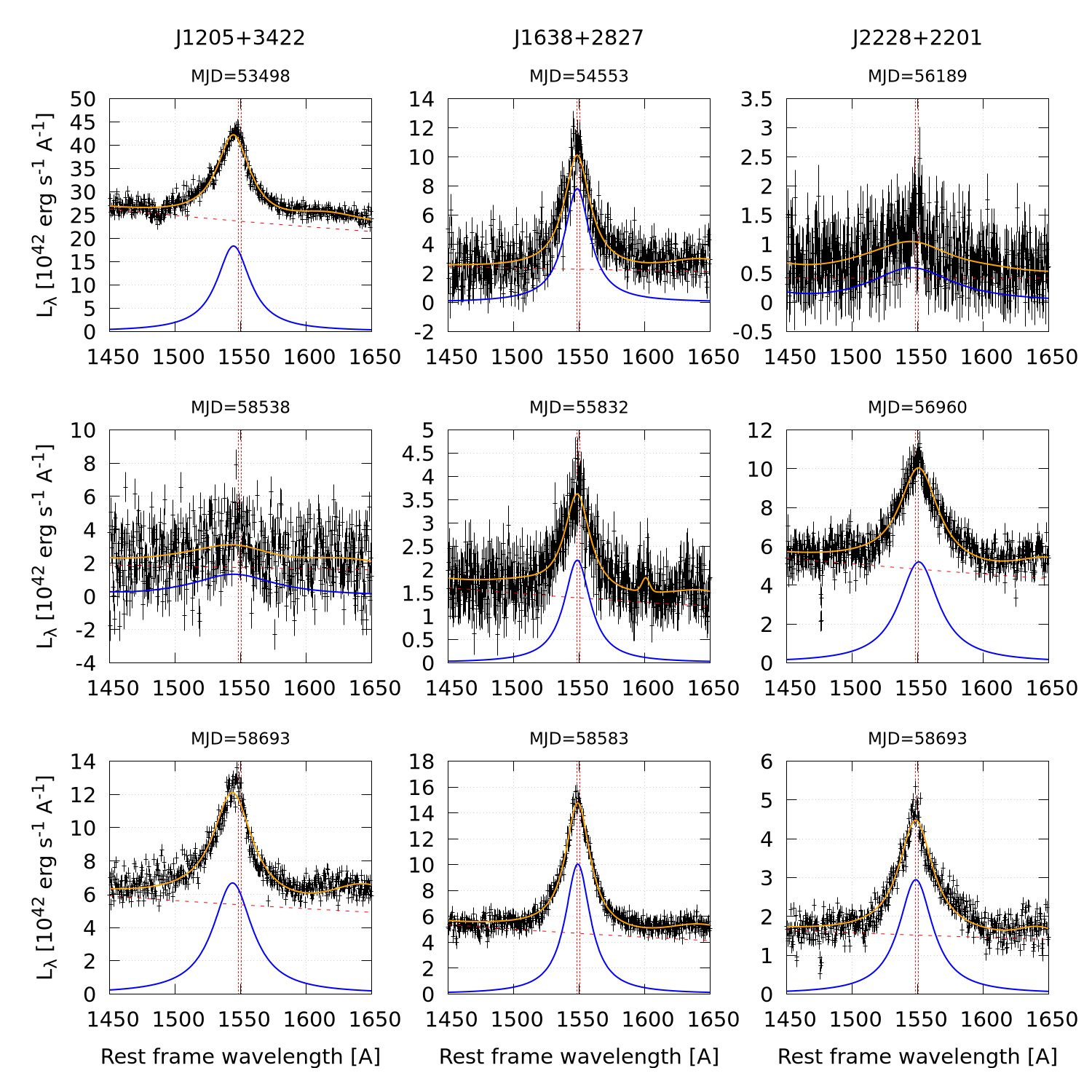}
    \vspace{-18pt}
  \caption{Observed spectra and best fit models of the \civ\ spectral region.
    The solid blue line is the Lorentzian profile fit to the \civ\ emission line.
    The long-dashed red line is the continuum fit, with the solid orange
    line giving the overall fit. The short-dashed vertical lines gives the
    rest wavelengths of the \civ\ doublet. }
  \label{fig:QSFit-CIV}
\end{figure}

\subsection{\civ\ Emission Line Evolution}
We analyzed the multi-epoch spectra of the three quasars using the
QSFit spectral fitting package \citep{Calderone2017}.  One advantage
of using QSFit is that it allows constraints on the slope and
luminosity of the broad band continuum of the source. The relevant
estimated quantities, including continuum luminosity and slope at
rest-frame 1450\AA, \civ\ line luminosity, FWHM, and EW are given in
Table~\ref{tab:QSFit-results}. All fits are performed with $E(B-V) =
0$ and the best fit models in the region of the \civ\ emission line
are shown in Figure~\ref{fig:QSFit-CIV}.

All \civ\ lines are fitted with a single, broad, Lorentzian profile.
This allows us to account for the central peak of the \civ\ line.  No
additional narrow components are considered for several reasons.
First, in the epochs of highest brightness, such a ``narrow''
component would have FWHM $\sim$ 2000-3000~km~s$^{-1}$, i.e. values
exceeding the usual widths of genuine narrow lines ($\lesssim
1000$~km~s$^{-1}$). Second, allowing a second component to have such
large widths would make that component highly degenerate with the
``broad'' components, causing the latter to have much larger widths,
$\sim 10,000$~km~s$^{-1}$.  Finally, neglecting such a narrow
component ensures a consistent model across all epochs.

Both the quasar continuum (evaluated at 1450\AA), and the \civ\ line
luminosities follow a similar evolution, with a ratio of $\sim$20-30,
confirming that the main driver for emission line variability is
likely the broad band continuum itself.  For all sources except
J1638+2827, the slope of the continuum changes with luminosity
following a ``bluer-when-brighter'' pattern, suggesting that a
distinct emerging component is responsible for both the slope and
luminosity variations.  In J1638+2827 the opposite behaviour is
observed, especially in the first observation epoch.  However, this
may be a bias due to the limited wavelength range available which
extends to rest-frame $\lambda \sim 1240$~\AA\ for the first epoch,
while it extend to shorter wavelengths for the other epochs
(respectively ~\AA\ and 1010~\AA).  This suggests that the emerging
component is more prominent at UV wavelengths, and a sufficient
wavelength coverage is required to detect it. In all cases where the
\civ\ line profile is reliably constrained, we find that the \civ\
FWHM is approximately constant with maximum variations $\lesssim 1000$
km s$^{-1}$, despite significantly larger line luminosity variations.

\begin{figure}
  \centering
  \includegraphics[width=8.5cm, trim=0.2cm 0.2cm 0.0cm 0.2cm, clip]
  {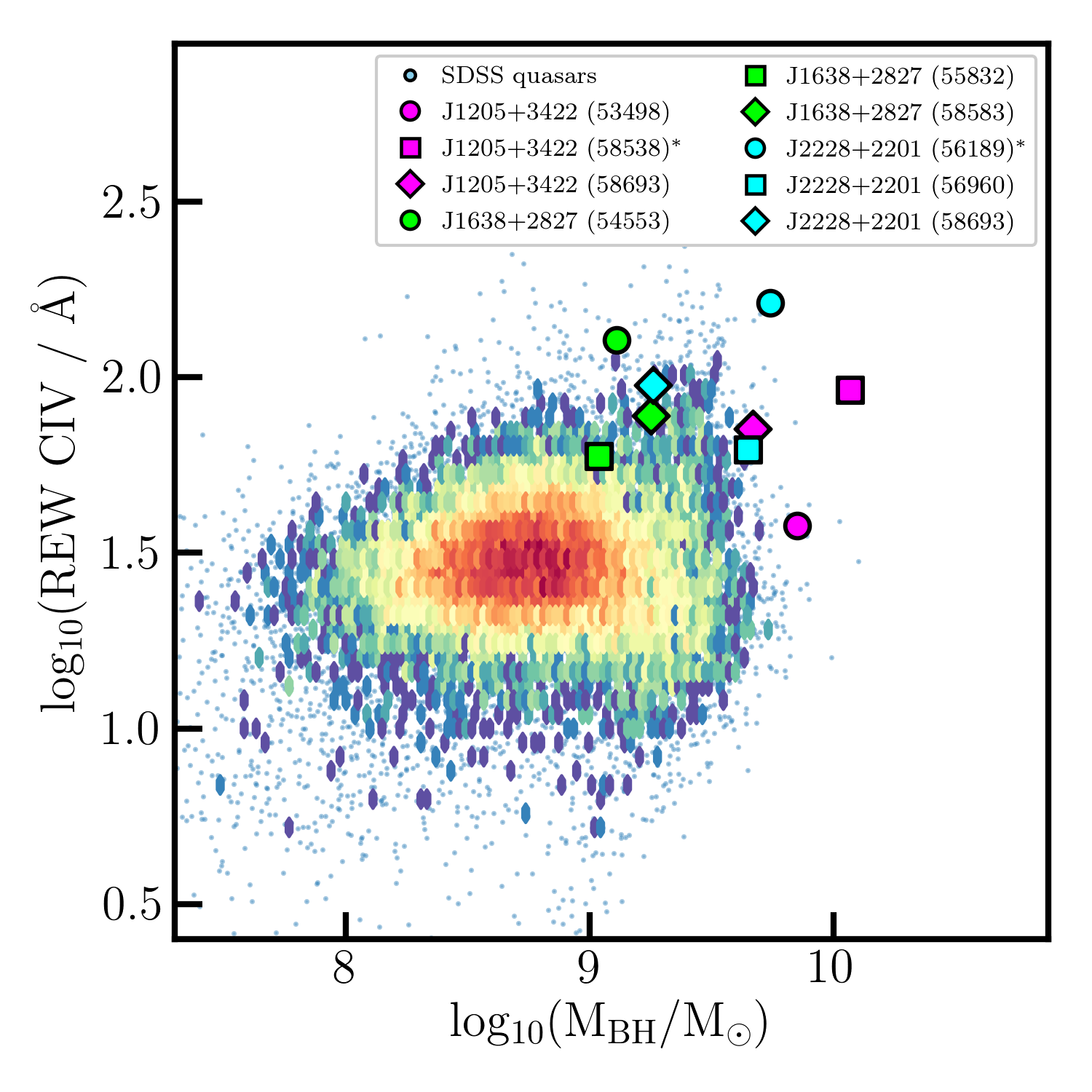}
   \vspace{-12pt}
   \caption[]{
     Colored density cloud shows virial black hole masses of
     $\approx$20,000 $z>1.5$ quasars from QSFit catalogue
     \citep{Calderone2017} compared to their \civ\ rest-frame equivalent
     widths (REWs).  Overplotted are the three high-redshift CLQs
     considered here: J1205+3422 (magenta), J1638+2827 (green), and
     J2228+2201 (cyan).  The symbol shapes indicate the observation epoch,
     with the first, second, and third epochs denoted by circles, squares,
     and diamonds, respectively. As per Table~\ref{tab:QSFit-results}, an
     asterisk in the inset box signifies that the \civ\ line for that
     quasar observation is very weak, likely making the associated
     estimates unreliable.}
   \label{fig:CIV_MBHvsREW}
\end{figure}

\subsection{Virial Black Hole Masses}
\label{sec:BH_masses} 
The FWHMs of quasar broad lines are related to the mass of the SMBH
powering the quasar phenomenon, and that mass is assumed to be
constant on any human timescale.  Hence it is instructive to check
whether the virial products, which are the basic quantity used to
calculate single-epoch black hole mass estimates, show any
variation. This approach assumes that the broad-line region (BLR) is
virialized \citep[e.g.][]{Shen2011, Calderone2017}. The continuum
luminosity provides a proxy for the BLR radius, and the broad-line
width (FWHM) provides a proxy for the virial velocity.  This ``virial
mass'' estimate is then expressed as:
\begin{equation}
  \log(M_{\rm BH}) = (\nu L_{\nu})^{\gamma} \times ({\rm FWHM})^{\delta}
\end{equation}
where $\gamma=0.5$ and $\delta=2$.  The virial product for the \civ\
CLQs is reported in the last column of Table~\ref{tab:QSFit-results}.
The uncertainty typically associated with single-epoch mass estimates
is $\sim 0.5$~dex, implying that the virial products for our three
high-redshift CLQs are all remarkably constant and compatible with
single black hole masses per source, even in those cases where the
\civ\ estimates were deemed potentially unreliable. The object showing
the largest variation is J2228+2201, though even that source only
spans a range of 0.48~dex.

\citet{Shen2011} and \citet{Kozlowski2017} report virial black hole
masses based on \mgii\ and \civ\ for the SDSS/BOSS observations of our
targets. For J1205+3422, \citet{Shen2011} report $\log (M_{\rm BH,
MgII}/M_{\odot}) = 9.55 \pm 0.05$ and $\log(M_{\rm BH, CIV}/M_{\odot})
= 9.49 \pm 0.04$. We use the mean of five values --- three from our
three epochs and two from \citet{Shen2011} --- to calculate an adopted
SMBH mass of $\log (M_{\rm BH, vir}/M_{\odot}) = 9.73$ for J1205+3422.
J1638+2827 also has two virial mass measurements from
\citet{Shen2011}, as well as two measurements from
\citet{Kozlowski2017}. The mean mass measurement adopted for
J1638+2827 is $\log(M_{\rm BH, vir}/M_{\odot} = 9.09$. For J2228+2201,
there are two \citet{Kozlowski2017} measurements, leading to a mean
adopted mass of $\log(M_{\rm BH, vir}/M_{\odot} = 9.37$. We use these
mean SMBH masses when calculating the Eddington ratios.

From the virial mass estimates, all our objects have SMBH masses
$M_{\rm BH} >10^{9} M_{\odot}$.  This is at the upper end of SMBH
masses at all epochs, and towards the extreme of the mass distribution
for $z\sim2$ quasars.  Figure~\ref{fig:CIV_MBHvsREW} compares the
\civ\ EW versus the virial SMBH masses for our multi-epoch
observations to a sample of $\approx 20,000 z>1.5$ SDSS quasars from
the QSFit catalog \citep{Calderone2017}.

\begin{figure}
  \centering
  \includegraphics[width=8.5cm, trim=0.2cm 0.2cm 0.0cm 0.2cm, clip]
  {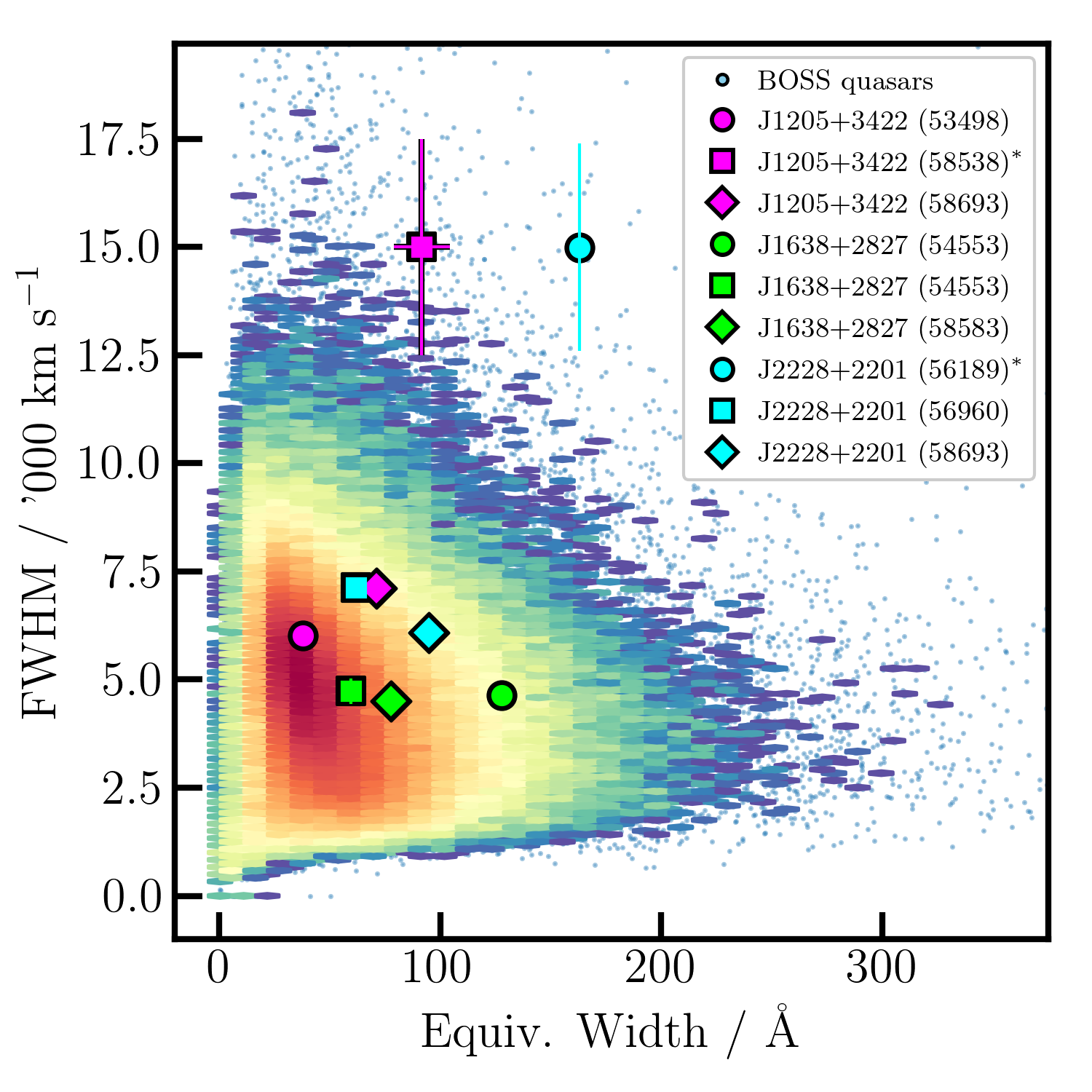}
  \vspace{-12pt}
  \caption[]{Rest-frame EW vs. FWHM of \civ\
    for BOSS DR12 quasar sample \citep{Hamann2017}.
   Symbols as in Figure~\ref{fig:CIV_MBHvsREW}.}
  \label{fig:REWvsFWHM}
\end{figure}

\subsection{Quantified Temporal Evolution of \civ\ Emission}
Quasars with interesting physical properties, such as extreme
outflows, can be selected using EW and FWHM measurements \citep[e.g.,
``Extremely Red Quasars'' ---][]{Ross2015, Zakamska2016, Zakamska2019,
Hamann2017}. Figure~\ref{fig:REWvsFWHM} shows the rest-frame EW versus
the FWHM of the \civ\ emission line in the BOSS DR12 quasar sample
from the catalogue of \citet{Hamann2017}.  Other than the two low
quality, suspect observations, the multi-epoch observations of our
high-redshift CLQs are all consistent with the bulk of the BOSS DR12
sample.

\begin{figure*}
  \centering
  \includegraphics[width=\textwidth]{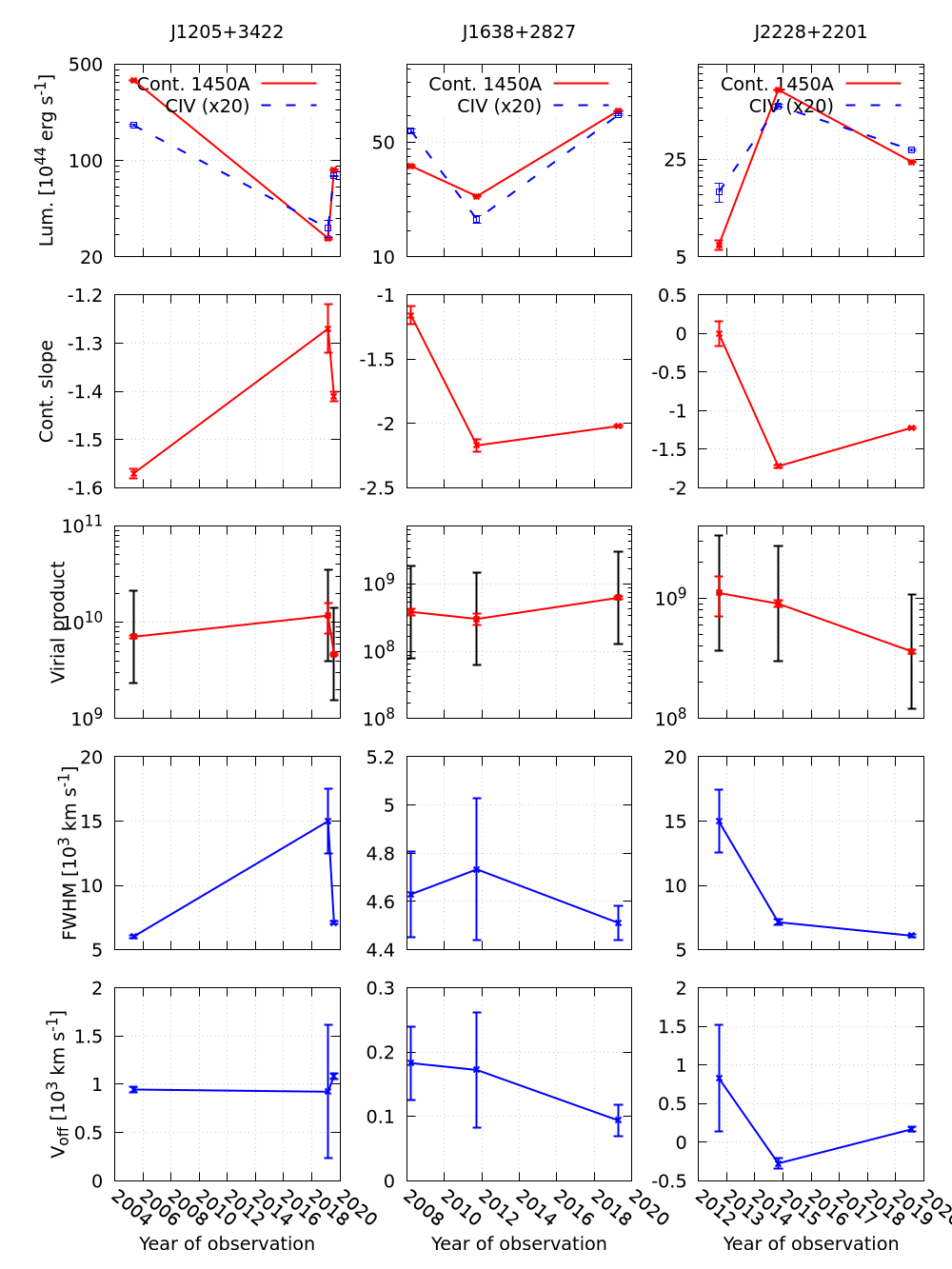}
  \vspace{-12pt}
  \caption{Temporal evolution of the spectral properties of the three
    quasars considered in this work.}
  \label{fig:QSFit-results}
\end{figure*}

The temporal evolution of the velocity offsets, 1450~\AA\ continuum
luminosity, continuum slope, \civ\ line width (FWHM), and virial black
hole mass estimates are shown in Figure~\ref{fig:QSFit-results}.  The
\civ\ velocity offsets are approximately constant, and generally
compatible with a single value (within 3$\sigma$).  The exception is
J2228+2201, where a significant change ($\sim$7$\sigma$) is observed
between the second and third epochs. The velocity blueshifts ($\sim
200-1100$ km s$^{-1}$) are consistent with rest-frame UV spectra of
quasars over the redshift range $1.5 \leq z \leq 7.5$
\citep[e.g.,][]{Meyer2019}.

\begin{table}
  \centering
  \begin{tabular}{l l l r}
    \hline
    \hline
    Object       & MJD & $L_{\rm bol}$&  $\eta_{\rm Edd}$  \\
    \hline
                     & 53498 & 47.216     &  $-$0.610   \\
 J1205+3422 & 58538 &  46.070    &   $-$1.756  \\   
                     & 58693 &  46.567    &  $-$1.259  \\                    
    \hline 
                     & 54553 & 46.166   & $-$1.025  \\
J1638+2827  & 55832 & 45.981   & $-$1.210  \\      
                     & 58583 & 46.504   & $-$0.687   \\
    \hline 
                     & 56189  & 46.231  &  $-$1.235 \\
J2228+2201  & 56960 & 47.342  & $-$0.124  \\      
                     & 58693 & 46.826   &  $-$0.640 \\
    \hline
    \hline
  \end{tabular}
  \caption{Bolometric luminosities are given in either \citet{Shen2011} or
    \citet{Kozlowski2017} for the three CLQs. We scale these values using
    our measured $\nu L_{\nu}$ continuum luminosity values.  From
    Section~\ref{sec:BH_masses}, we assume: $\log(M_{\rm BH,
      vir}/M_{\odot}) = 9.726$ for J1205+3422; $\log(M_{\rm BH,
      vir}/M_{\odot}) = 9.091$ for J1638+2827, and $\log(M_{\rm BH,
      vir}/M_{\odot}) = 9.366$ for J2228+2201.  The Eddington luminosity
    $L_{\rm Edd} = 1.26\times10^{38} (M_{\rm BH}/M_{\odot})$ erg s$^{-1}$
    and $\eta_{\rm Edd}$ is the log of the Eddington ratio.}
\label{tab:Eddington_ratios} 
\end{table}

\begin{figure}
  \centering
  \includegraphics[width=8.5cm, trim=0.2cm 0.2cm 0.0cm 0.2cm, clip]
  {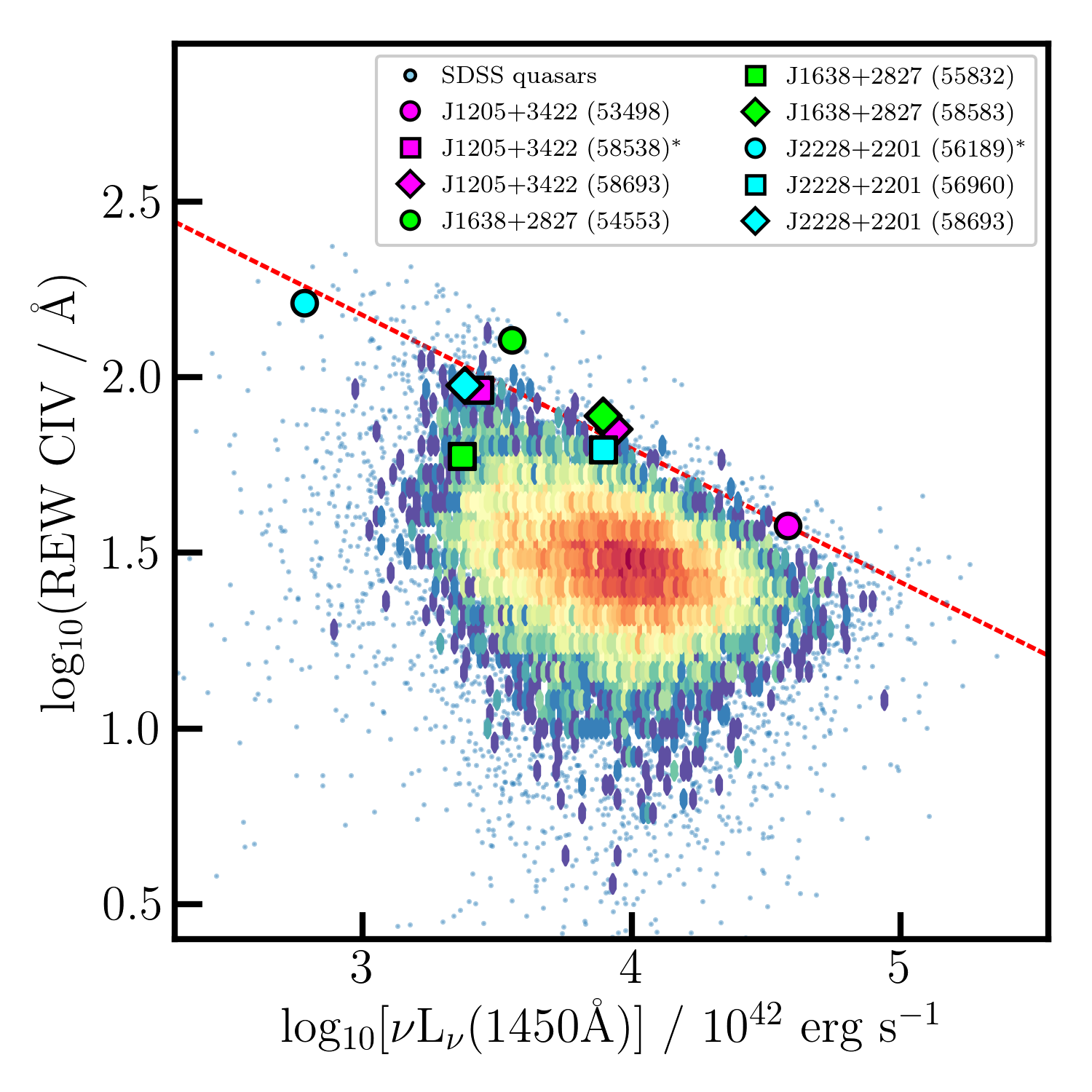}
   \vspace{-12pt}
   \caption[]{\civ\ equivalent width versus underlying continuum luminosity,
     commonly referred to as the Baldwin plot.  The continuum luminosities
     are from \citet{Calderone2017}.  The rest-frame EW (REW) measurements
     are from Table~\ref{tab:QSFit-results}.  Symbols as in
     Figure~\ref{fig:CIV_MBHvsREW}.  The dashed red line has slope
     $\beta=-0.38$.}
  \label{fig:CIV_Baldwin}
\end{figure}

\subsection{The \civ\ Baldwin Effect}
The Baldwin effect \citep{Baldwin1977} is an empirical relation
between quasar emission line rest-frame EWs and continuum luminosity
\citep[e.g.,][]{Shields2007, Hamann2017, Calderone2017}.
\citet{Hamann2017} and \citet{Calderone2017} present recent
measurements of the Baldwin effect for large quasar samples.

There is an anti-correlation between the rest-frame EWs and, e.g.,
1450~\AA\ rest-frame continuum luminosity, so that as the underlying
UV continuum luminosity increases, the EW decreases.
Figure~\ref{fig:CIV_Baldwin} shows this for a sample of 20,374 quasars
from the QSFit catalogue \citep{Calderone2017}.  The slope is $\beta$
is $-0.20$, consistent with \citet[][ $\beta=-0.25$, but for
bolometric luminosity rather than rest-frame UV continuum luminosity]
{Kozlowski2017} and with \citet[][ $\beta=-0.23$]{Hamann2017}.

We add the measurements from the three \civ\ CLQ quasars at each epoch
to Figure~\ref{fig:CIV_Baldwin}. All three quasars at all three epochs
lie on the edge of the $\nu L_{\nu}$-EW distribution.  Also, with the
exception of J1638+2827 on MJD 55832, all the measurements show an
{\it intrinsic} Baldwin effect \citep[e.g.,][]{Goad2004, Rakic2017}.
Subject to small number statistics, the slope of the intrinsic Baldwin
effect for these three high-redshift \civ\ CLQs is somewhat steeper
than the population trends, with $\beta \approx-0.38$, shown by the
dashed red line in Figure~\ref{fig:CIV_Baldwin}.

\section{Discussion}
\subsection{Continuum and Line Changes: Comparisons to recent Observations}
The top row of Figure~\ref{fig:QSFit-results} demonstrates that both
the 1450~\AA\ continuum and the \civ\ emission lines can exhibit
large, greater than an order-of-magnitude, changes in luminosity, {\it
and} that these continuum-line changes track each other.

\citet{Trakhtenbrot2019} report on the quasar 1ES~1927+654 at
$z=0.02$, which was initially seen to lack broad emission lines and
line-of-sight obscuration, i.e, it was a ``type 2'' quasar. This
object was then seen to spectroscopically change with the appearance
of a blue, featureless continuum, followed by the emergence of broad
Balmer emission lines --- i.e., this quasar changed into a broad-line,
or ``type 1'' quasar, after the continuum luminosity brightened.  This
implies that there is (at least in some cases) a direct relationship
between the continuum and broad emission lines in CLQs.  Many other
examples of a similar phenomenon have been reported, albeit not with
the dense spectral cadencing of 1ES~1927+654.  A similar scenario may
have occurred for the three high-redshift CLQs presented here,
although we lack the high-cadence, multiwavelength, multi-epoch
coverage that \citet{Trakhtenbrot2019} present.  Interestingly,
\citet{Trakhtenbrot2019} find no evidence for broad UV emission lines
in their source, including neither \civ, \ciii, nor \mgii.  The
authors attribute the lack of broad UV emission lines to dust within
the BLR.

\citet{MacLeod2019} present a sample of CLQs where the primary
selection requires large amplitude ($| \Delta g | > 1$ mag, $| \Delta
r | > 0.5$ mag) variability over any of the available time baselines
probed by the SDSS and Pan-STARRS1 surveys. They find 17 new CLQs,
corresponding to $\sim 20$\% of their observed sample. This CLQ
fraction increases from 10\% to roughly half, as the rest-frame
3420~\AA\ continuum flux ratio between repeat spectra increases from
1.5 to 6 --- i.e., more variable quasars are more likely to exhibit
large changes in their broad emission lines. \citet{MacLeod2019} note
that their CLQs have lower Eddington ratios relative to the overall
quasar population.

Using the same dataset as \citet{MacLeod2019}, \citet{Homan2019}
investigate the responsiveness of the \mgii\ broad emission line
doublet in AGN on timescales of several years.  Again focussing on
quasars that show large changes in their optical light-curves,
\citet{Homan2019} find that \mgii\ clearly does respond to the
continuum.  However, a key finding from \citet{Homan2019} is that the
degree of responsivity varies strikingly from one object to another.
There are cases of \mgii\ changing by as much as the continuum, more
than the continuum, or very little at all.  In the majority (72\%) of
this highly variable sample, the behaviour of \mgii\ corresponds with
that of H$\beta$.  However, there are also examples of \mgii\ showing
variations while but H$\beta$ does not, and vice versa.

\citet{Graham2019b} report the largest number of H$\beta$ CLQs to
date, with a sample of 111 sources identified. \citet{Graham2019b}
find that this population of extreme varying quasars, referred to as
CSQs in their paper, is associated with {\it changes} in the Eddington
ratio rather than simply the magnitude of the Eddington ratio itself.
They also find that the timescales imply cooling/heating fronts
propagating through the disk.

\subsection{Continuum and Line Changes: Comparisons to Theoretical Expectations}\label{sec:theory}
The \civ\ line is one of the strongest collisionally excited lines in
quasar spectra \citep[e.g.,][]{HamannFerland1999}, and \civ\ emission
probes the photoionisation environment produced by the innermost disk,
as indicated by reverberation mapping time-delay measurements.

In standard \citet{SS73} thin disk models, large changes in the
continuum flux are not permitted over short timescales due to the
relatively long viscous time associated with such disks. Given the
observed short timescale continuum variations, we must consider more
complex models. We note that two of our sources (J1205+3422 and
J2228+2201) have \civ\ EW and continuum luminosity changes which fall
comfortably along a line, implying their variation is consistent with
an intrinsic Baldwin effect (see Figure~\ref{fig:CIV_Baldwin}). That
is, the excitation of the \civ\ line is driven by the changes in the
continuum; however, the excited gas cannot reach an equilibrium
excitation state as quickly as the continuum changes, making the slope
of the intrinsic Baldwin effect greater than that of the overall
ensemble slope, which is derived from many single-epoch observations,
the majority of which are assumed to be in equilibrium.

The presence of an intrinsic Baldwin effect implies those sources may
comfortably fit into the sample of \civ\ variable quasars explored by
\citet{Dyer2019}. Similar to those authors, we consider slim accretion
disk models \citep[e.g.,][]{Abramowicz1988}, which may explain the
observed variability. In particular, the summary of disk variation
timescales presented in \citet{Stern2018} shows that timescales are
shorter for taller disks, permitting changes similar to those observed
if they are caused by heating or cooling fronts propagating on the
disk sound crossing time. \citet{Dyer2019} also consider inhomogeneous
disk models \citep[e.g.,][]{DexterAgol2011} where flux variations are
driven by azimuthal inhomogeneities in the temperature of the
disk. Such inhomogeneities could arise from disk instabilities
\citep[e.g.,][]{LightmanEardley1974} or interactions of embedded
objects \citep[e.g.,][]{McKernan2014, McKernan2018}. If an
inhomogeneous disk is responsible for the continuum variations, which
then drive \civ\ variations, it implies that our objects are simply
the extreme outliers produced by a process which occasionally, but
rarely, produces very large hot or cool spots. The frequency of such
occurrences can be used to constrain the slope and normalization of
the power law distribution of spot size.

Finally, we must also consider the disk/wind model which has
successfully reproduced many features of quasar \civ\ observations,
notably the common blueshifted offset \citep[e.g.,][]{Murray1995}. In
this model, \civ\ is optically thick at low velocities, but optically
thin at high velocities, with $\tau \sim 1$ at $\sim$5000 km
s$^{-1}$. In a disk/wind model, the BLR is very small, implying short
timescales of emission line variability. Also in this model, there
should be associated strong absorption in the soft X-ray band, which
could be tested with follow-up observations.

The foregoing discussion directly applies to the two objects which
maintain an intrinsic Baldwin effect relationship between \civ\ EW and
continuum luminosity. However, J1638+2827 clearly does not maintain
such a relationship --- indeed, the continuum and \civ\ EW appear
somewhat correlated, rather than anti-correlated, in this source. This
implies that the illumination of the \civ\ emitting region by a
variable ionising continuum and the corresponding change in the
photoionisation state alone cannot explain the collapse and recovery
of the line.

\begin{figure*}
  \centering
  \includegraphics[width=14.5cm, trim=0.2cm 0.2cm 0.0cm 0.2cm, clip]
  {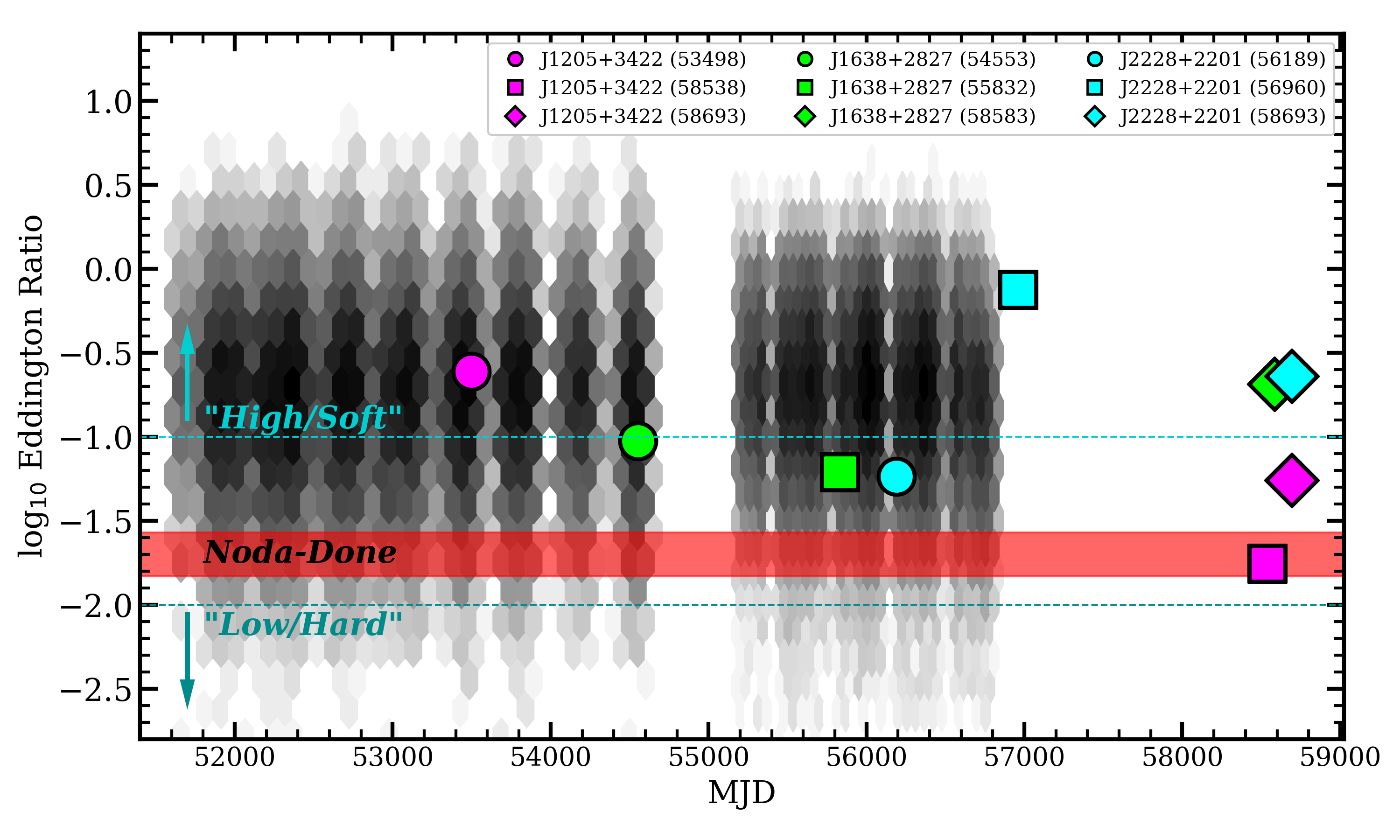}
  \vspace{-12pt}
  \caption[]{Eddington ratios of the three \civ\ CLQs.  Grayscale gives the
    Eddington ratio ranges for quasars from SDSS \citet[][]{Shen2011} and
    BOSS \citep[][]{Kozlowski2017}.  Symbols as in
    Figure~\ref{fig:CIV_MBHvsREW}.  The red region ($L / L_{\rm Edd}
    \approx 0.02 \pm 0.01$; $\eta_{\rm Edd} = -1.7 \pm 0.13$) is the
    transition accretion rate suggested by \citet{NodaDone2018}.  A
    ``High/Soft State'' with $\eta_{\rm Edd} \geq -1$ and a a ``Low/Hard
    State'' with $\eta_{\rm Edd} \leq -2$ are indicated with dashed lines
    and arrows.}
  \label{fig:Eddington_ratios}
\end{figure*}

\subsection{Eddington Ratios and State Changes} 
The broad UV and optical lines in quasars are most sensitive to the
extreme ultraviolet (EUV) part of the SED, with \civ\ (and indeed
\heii\ and N\,{\sc v}) being at the higher energy end of the EUV
distribution.

\citet{NodaDone2018} note that the soft X-ray excess -- the excess of
X-rays below 2 keV with respect to the extrapolation of the hard X-ray
spectral continuum model, and a common feature among Type~1 AGN --
produces most of the ionizing photons, so its dramatic drop can lead
to the disappearance of the broad-line region, driving the
changing-look phenomena. \citet{NodaDone2018} go on to make a
connection between state changes in Galactic binaries and CLQs.
However, one thing to note is that radiation pressure should be much
more important in AGNs than in stellar mass black hole systems,
implying that sound speed is much faster than expected from the gas
temperature alone.  In Galactic binaries spectral hardening
corresponds to the soft-to-hard state transition in black hole
binaries, which occurs at $L / L_{\rm Edd} \sim 0.02$ (i.e.,
$\eta_{\rm Edd} \sim -1.7$).  During this state transition, the inner
disc evaporates into an advection-dominated accretion flow, while the
overall drop in luminosity appears consistent with the hydrogen
ionisation disc instability.  \citet{NodaDone2018} predict that
changing-look AGNs are similarly associated with state transitions at
$L / L_{\rm Edd}$ of a few percent.

Comparing observed correlations between the optical-to-X-ray spectral
index (i.e., $\alpha_{\rm ox}$) and Eddington ratio in AGN to those
predicted from observations of X-ray binary outbursts,
\citet{Ruan2019a} find a remarkable similarity to accretion state
transitions in prototypical X-ray binary outbursts, including an
inversion of this correlation at a critical Eddington ratio of
$\sim$10$^{-2}$, i.e., at the same ratio motivated by
\citet{NodaDone2018}.  These results suggest that the structures of
black hole accretion flows directly scale across a factor of
$\sim10^{8}$ in black hole mass and across different accretion
states. Using \citet{Ruan2019a} as a guide, there are potentially
three accretion regimes: {\it (i)} a ``High/Soft State'' with
$\eta_{\rm Edd} \gtrsim -1$; {\it (ii)} a ``Low/Hard State'' with $-2
\lesssim \eta_{\rm Edd} \lesssim -1$; and {\it (iii)} a ``Low/Hard
State'' with $\eta_{\rm Edd} \lesssim -2$.  These are given as shaded
regions in Figure~\ref{fig:Eddington_ratios}.

Apart from J1205+3422 on MJD 58538, we see all the \civ\ CLQ epochs
are above $L / L_{\rm Edd} \sim 5$\%. As we have noted throughout, the
spectrum for J1205+3422 on MJD 58538 has low SNR, so while this object
could well have entered a `low-state', this is difficult to
conclusively confirm.  Nevertheless, and taking the J1205+3422
spectrum at face-value, we note that all the \civ\ CLQs are well above
$\sim1\%$ in Eddington luminosity and thus the suggested ``Low/Hard
State'' boundary.

While we note there is interest in plotting these accretion rates, and
motivation from \citet{NodaDone2018} and \citet{Ruan2019a}, we very
much caution on over-interpretation at this juncture. The point of
this paper was to report on three very interesting CLQs.  Our object
definitions are based on empirical, observed properties.  The
potential for physical connections of accretion physics across the
large-range of mass-scales is tantalising, but is left to future
investigations.

\section{Conclusions}
In this paper we have reported on three redshift $z>2$ quasars with
dramatic changes in their \civ\ emission lines, the first sample of
changing-look quasars at high redshift.  This is also the first time
the changing-look behaviour has been seen in a high-ionisation
emission line.

\begin{itemize}
\item SDSS J1205+3422, J1638+2827 and J2228+2201 show interesting
  behaviour in their observed optical light curves, and subsequent
  spectroscopy shows significant changes in the \civ\ broad emission
  line, with both line collapse and emergence being displayed on
  rest-frame timescales of $\sim 240-1640$~days.
\item Where observed, the profile of the Ly$\alpha$/\nv emission complex
  also changes, and there is tentative evidence for changes in the \mgii\
  line.
\item Although line measurements from the three quasars show large changes
  in the \civ\ line flux-line width plane, the quasars are not seen to
  be outliers when considered against the full quasar population
  in terms of (rest-frame) EW and FWHM properties.
\item 
  We put these observations in context with recent ``state-change''
  models, and note that with the exception of the low SNR spectrum of
  J1205+3422 on MJD 58538, even in their low state, 
  the \civ\ CLQs are above $\sim5$\% in Eddington luminosity. 
\end{itemize}

There are now more observed examples of dramatic, dynamic changes in
supermassive black hole systems e.g. CLQs/CSQs than Galactic X-ray
Binaries.  While the timescales are expected to scale with black hole
mass, whether the full physical system at large does, including the
associated atomic physics, remains an open question.

\subsection*{Availability of Data and computer analysis codes} 
All materials, databases, data tables and code are fully available at: 
\href{https://github.com/d80b2t/CIV_CLQs}{\tt https://github.com/d80b2t/CIV\_CLQs}.

\section*{Acknowledgements}
We thank Andy Lawrence, Mike Hawkins and David Homan for useful discussion.
NPR acknowledges support from the STFC and the Ernest Rutherford Fellowship scheme. 
MJG is supported in part by the NSF grants AST-1815034, and the NASA grant 16-ADAP16-0232.
\\

This paper heavily used \href{http://www.star.bris.ac.uk/~mbt/topcat/}{TOPCAT} (v4.4)
\citep[][]{Taylor2005, Taylor2011}.
This research made use of \href{http://www.astropy.org}{\tt Astropy}, 
a community-developed core Python package for Astronomy 
\citep{AstropyCollaboration2013, AstropyCollaboration2018}.

Funding for SDSS-III has been provided by the Alfred P. Sloan
Foundation, the Participating Institutions, the National Science
Foundation, and the U.S. Department of Energy Office of Science. The
SDSS-III web site is
\href{http://www.sdss3.org/}{http://www.sdss3.org/}.
SDSS-III is managed by the Astrophysical Research Consortium for the
Participating Institutions of the SDSS-III Collaboration including the
University of Arizona, the Brazilian Participation Group, Brookhaven
National Laboratory, Carnegie Mellon University, University of
Florida, the French Participation Group, the German Participation
Group, Harvard University, the Instituto de Astrofisica de Canarias,
the Michigan State/Notre Dame/JINA Participation Group, Johns Hopkins
University, Lawrence Berkeley National Laboratory, Max Planck
Institute for Astrophysics, Max Planck Institute for Extraterrestrial
Physics, New Mexico State University, New York University, Ohio State
University, Pennsylvania State University, University of Portsmouth,
Princeton University, the Spanish Participation Group, University of
Tokyo, University of Utah, Vanderbilt University, University of
Virginia, University of Washington, and Yale University.

This publication makes use of data products from the {\it Wide-field
Infrared Survey Explorer}, which is a joint project of the University
of California, Los Angeles, and the Jet Propulsion
Laboratory/California Institute of Technology, and {\it NEOWISE}, which is a
project of the Jet Propulsion Laboratory/California Institute of
Technology. {\it WISE} and {\it NEOWISE} are funded by the National Aeronautics
and Space Administration.

Based on observations obtained with the Samuel Oschin 48-inch Telescope at the Palomar Observatory as part of the Zwicky Transient Facility project. ZTF is supported by the National Science Foundation under Grant No. AST-1440341 and a collaboration including Caltech, IPAC, the Weizmann Institute for Science, the Oskar Klein Center at Stockholm University, the University of Maryland, the University of Washington, Deutsches Elektronen-Synchrotron and Humboldt University, Los Alamos National Laboratories, the TANGO Consortium of Taiwan, the University of Wisconsin at Milwaukee, and Lawrence Berkeley National Laboratories. Operations are conducted by COO, IPAC, and UW.

\bibliographystyle{mnras}
\bibliography{CIV_CLQs}

\bsp	
\label{lastpage}
\end{document}